\documentclass[twocolumn]{aastex61}
\pdfoutput=1 
\usepackage{amsmath,amstext}
\usepackage[T1]{fontenc}
\usepackage{apjfonts} 
\usepackage[figure,figure*]{hypcap}

\def\arcsec{\hbox{$^{\prime\prime}$}}

\def\farcm{\hbox{$.\mkern-4mu^\prime$}}

\shorttitle{AASTeX 6.1 Template}
\shortauthors{Beccari et al.}

\begin{document}

\title{Discovery of a double Blue Straggler sequence in M15: new insight into the core-collapse process}

\author{G. Beccari}
\affiliation{European Southern Observatory, Karl-Schwarzschild-Strasse 2, 85748  Garching bei M\"unchen, Germany}
\author{F.R. Ferraro}
\affiliation{Dipartimento di Fisica e Astronomia, Universita di Bologna, Via Gobetti 93/2, Bologna, Italy}
\author{E. Dalessandro}
\affiliation{INAF Osservatorio di Astrofisica e Scienza dello Spazio di Bologna, Via Gobetti 93/3, Bologna, Italy}
\author{B. Lanzoni}
\affiliation{Dipartimento di Fisica e Astronomia, Universita di Bologna, Via Gobetti 93/2, Bologna, Italy}
\author{S. Raso}
\affiliation{Dipartimento di Fisica e Astronomia, Universita di Bologna, Via Gobetti 93/2, Bologna, Italy}
\affiliation{INAF Osservatorio di Astrofisica e Scienza dello Spazio di Bologna, Via Gobetti 93/3, Bologna, Italy}
\author{L. Origlia}
\affiliation{INAF Osservatorio di Astrofisica e Scienza dello Spazio di Bologna, Via Gobetti 93/3, Bologna, Italy}
\author{E. Vesperini}
\affiliation{Department of Astronomy, Indiana University, Swain West, 727 E. 3rd Street, IN 47405 Bloomington, USA}
\author{J. Hong}
\affiliation{Kavli Institute for Astronomy and Astrophysics, Peking University, Yi He Yuan Lu 5, HaiDian District, Beijing 100871, People's Republic of China}
\author{A. Sills}
\affiliation{Department of Physics and Astronomy, McMaster University, Hamilton, Ontario L8S 4M1, Canada}
\author{A. Dieball}
\affiliation{Physics and Astronomy, University of Southampton, Southampton SO17 1BJ, UK}
\affiliation{Helmholtz-Institut f\"ur Strahlen und Kernphysik, Nussallee 14-16, D-53115 Bonn, Germany}
\author{C. Knigge}
\affiliation{Physics and Astronomy, University of Southampton, Southampton SO17 1BJ, UK}

\begin{abstract}
In this paper we  report on the discovery of a double blue straggler star (BSS) sequence in the core of the core-collapsed cluster M15 (NGC 7078). We performed a detailed photometric analysis of the extremely dense core of the cluster using a set of images secured
with the Advanced Camera for Survey in the High Resolution Channel mode on-board the Hubble Space Telescope.  The proper combination of the large number of single frames in the near-UV  (F220W), and blue (F435W) filters allowed us to perform  a superb modeling of the Point Spread Function 
and an accurate deblending procedure. The Color-Magnitude diagram revealed the presence of two distinct parallel sequences of blue stragglers. 
In particular, the blue BSS sequence is characterized by the intriguing presence of two different branches. The first branch appears extremely narrow, it extends up to 2.5 magnitudes brighter than the cluster main-sequence turnoff (MS-TO) point, and it is nicely reproduced by a 2 Gyr-old collisional isochrone. The  second branch extends up to 1.5 magnitudes from the MS-TO and it is reproduced by a 5.5 Gyr-old collisional isochrone. Our observations suggest that each of these branches is mainly constituted by a population of nearly coeval collisional BSS of different masses generated during two episodes of high collisional activity. We discuss the possibility that the oldest episode corresponds to the core-collapse event (occurred about 5.5 Gyr ago), while the most recent one (occurred about 2 Gyr ago) is associated with a core oscillation in the post-core collapse evolution. The discovery of these features provides further strong evidence in support of the connection between the BSS properties and GC dynamical evolution, and it opens new perspectives on the study of core-collapse and post core-collapse evolution.

\end{abstract}


\section{Introduction}
In the color-magnitude diagram (CMD) of any stellar aggregates Blue straggler stars (BSSs) typically appear as a sparse and scattered population of stars lying in a region at brighter magnitudes and bluer colors with respect to the cluster Main Sequence turnoff (MS-TO) stars~\citep{sandage53, fe97, fe03,piotto04,la07,leigh07,fe12,simu16,fe18}.

Their location in the CMD, combined with further observational evidence \citep{shara97,gilliland98,fe06,fiore14}, suggests that they are hydrogen-burning stars more massive than the parent cluster stars at the MS-TO. These exotica have been suggested to be powerful tools to investigate the internal dynamical evolution of the parent cluster \citep{fe12,ale16,la16,fe18}. BSSs are generated through two main formation channels: (1) mass-transfer (MT) phenomena and/or coalescence in primordial binary systems \citep{mc64}, and (2) direct stellar collisions \citep{hills76}.  While the first process (which is driven by the secular evolution of binary systems) is common to any stellar environment (low and high-density stellar systems, as globular and open clusters, dwarf galaxies, the Galactic field), the second one requires a high-density environment, where the rate of  stellar collisions is increased. Note that a collisional environment can affect also the efficiency of the MT formation channel, since dynamical interactions involve single and binary stars thus favoring the tightening of binaries.  The congested central regions of high-density globular clusters (GCs) are the ideal habitat where collisions can occur. Hence in such an environment both the BSS channels can be active \citep{fe93,fe95}, although the MT formation channel seems to be the most active one (see \citealt{davies04,knigge09}).

Although a few intriguing spectroscopic features have been interpreted as the fingerprints of the MT formation channels (see \citealt{fe06}), BSSs produced by different formation processes still appear photometrically indistinguishable. In this respect, a promising route of investigation was opened by the discovery of two distinct  sequences of BSSs in the post core-collapse cluster M30 by \citet{fe09}. After this discovery, a similar feature has been detected in NGC 362 \citep{da13} and NGC~1261 \citep{si14}. In only one case has a double BSS sequence been detected in a young cluster in the Large Magellanic Cloud \citep{li18a,li18b}. However, it has been argued \citep{dal19a,dal19b} that the observed bifurcation could be an artifact due to field star contamination.

The main characteristic ubiquitously detected in these clusters is the presence of a narrow BSS sequence on the blue side of the CMD, separated through a clear-cut gap from a more scattered population of BSSs on the red side. The narrowness of the blue sequence demonstrates that it is populated by a nearly coeval population of stars with different masses, generated over a relatively short period of time. Moreover, the perfect agreement of the blue sequence locus with collisional isochrones \citep{sills09} suggests that these objects have been formed through collisions. Instead, the red sequence is by far too red to be reproduced by collisional models of any age. All these facts support the hypothesis that the origin of the blue BSS sequence is related to a phenomenon able to enhance the probability of collisions over a relatively short period of time, thus promoting the formation of collisional BSSs. \citet{fe09} proposed that this phenomenon is the cluster core-collapse (CC). This hypothesis has been recently confirmed by numerical simulations \citep{pz19}
\begin{figure*}
\centering
\includegraphics[width=11.9truecm]{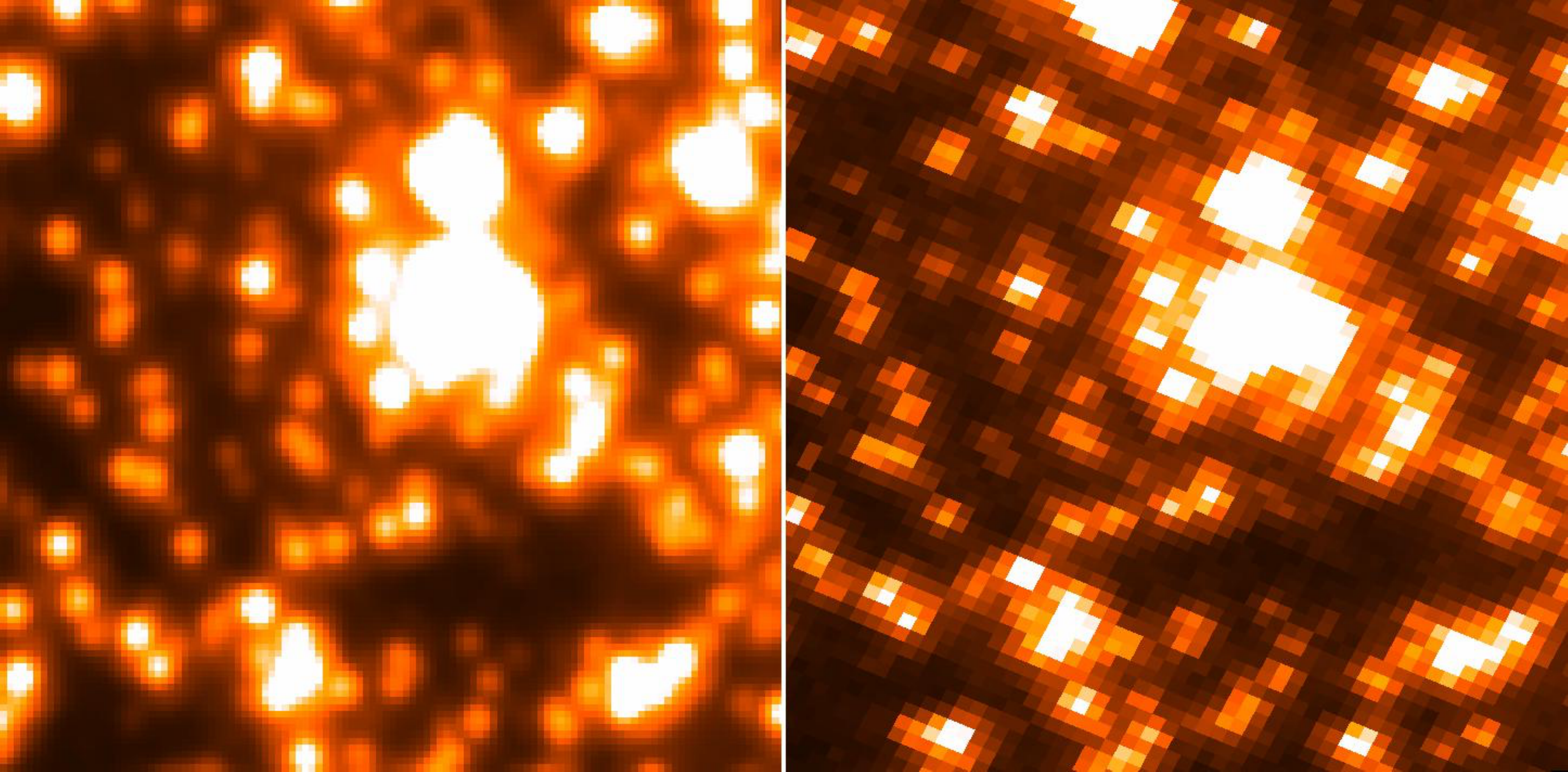}
\caption{Comparison between the central $1.5\arcsec\times1.5\arcsec$ region of M15 as seen through the super sampled B$_{435}$ image, and that observed through a single exposure in the same filter (left and right panels, respectively). North is up and East is left.}
\label{fig:frame}
\end{figure*}
%

In this context, \citet{fe09} also suggested that the properties of the blue-BSS sequence (essentially its extension in luminosity) can be used to date the epoch of CC. In short, while the MS-TO luminosity is a sensitive indicator of the age of the cluster, the luminosity of TO point of the blue BSS sequence can be used to determine the epoch at which the collisional event has promoted the formation of this stellar population.
In fact the comparison with collisional models allowed the dating of the epoch of the CC event: approximately 2 Gyr ago in M30 \citep{fe09} and NGC 1261 \citep{si14}, and 0.2 Gyr ago in NGC 362 \citep{da13}.

Recent Monte-Carlo simulations of synthetic MT-BSSs  \citep{ji17} showed that the blue-BSS sequence can be also populated by MT-BSSs.  This finding by itself does not invalidate the possibility that the blue BSS sequence might be mainly populated by collisional BSSs, since the easiest way to reproduce the observed narrowness of the blue sequence and the adjacent gap is still the hypothesis that the vast majority of BSSs along this sequence formed almost simultaneously, instead as a result of a continuous process extending over a long time interval. Conversely, the result of \citet{ji17} might explain the presence of the few W Uma stars (which are close binaries) actually detected by \citet{fe09} and \citet{da13} along the blue BSS sequence in M30 and in NGC 362. In this framework it is important to stress that, although the location in the CMD  cannot be used to  individually distinguish collisional from MT-BSSs, it mirrors the occurrence of the two formation mechanisms: when a narrow blue sequence is detected, this is mainly populated by collisional-BSSs, with some possible contamination from MT-BSSs (as the WUma variables detected along the blue sequences); the red side, instead, cannot be reproduced by any of the available collisional models \citep{sills09} and is mainly populated by  MT-BSSs with some contamination from evolved BSSs generated by collisions.

  
In this paper we present the discovery of a double BSS sequence in the post-core collapse cluser M15 (NGC7078). Indeed, M15 is one of the most massive ($log M = 6.3M_{\odot}$, several $10^6 M_{\odot}$) and dense ($log \rho_0 =6.2 M_{\odot}/pc^3$) cluster of the Galaxy. As emphasised by~\citet[][]{le13}, while binary evolution seems to be the dominant channel for BSS formation, the fact that M15 has one of the densest and most massive cores among Galactic GCs, makes it the ideal environment where to expect the collisional formation channel to be relevant.
This fact has been recently confirmed by~\citet[][]{le19} where the authors foresee the collisions via 2+2 interactions to play a crucial role on the formation of BSSs in post-core collapse clusters. 
Since the very first detections of a central cusp in its surface brightness profile \citep{dk86,lauer91,ste94} it has been considered as a sort  of ``prototype" of post-CC clusters, and it is still catalogued so in all the most recent GC compilations (see the updated version of the Harris Catalog). The BSS sequence studied here shows a complex structure that might provide crucial information on the core-collapse phenomenon.

\section{Data reduction}
\label{sec_data}
 
In this work we use a set of archival high resolution images obtained with the Advanced Camera for Survey (ACS) on board the Hubble Space Telescope (HST). The core of the cluster was observed with the High Resolution Channel (HRC) mode of the ACS in the $F220W$ and $F435W$ filters (hereafter, U$_{220}$ and B$_{435}$, respectively). The HRC provides a supreme spatial resolution, with a pixel scale of $\sim0.028\times0.025$ arcsec$^2$/pixel in a field of view of $29\arcsec\times26\arcsec$. A total of 13 images of 125s of exposure time each were acquired in the B$_{435}$ filter under the program GO-10401 (PI Chandar), while 
8$\times$290s exposures were taken with the U$_{220}$ filter under the program GO-9792 (PI Knigge). These images were already used in~\citet[][]{die07} and~\citet[][]{ha10} to probe the stellar population in the innermost regions of the cluster. As described in~\citet[][]{ha10}, the HRC images were combined using a dedicated software developed by Jay Anderson and similar to DrizzlePack\footnote{http://www.stsci.edu/hst/HST\_overview/drizzlepac}. In short, the software combines single raw images of a given band to improve the sampling of the Point Spread Function (PSF). 
A $3\arcsec\times3\arcsec$ section of a super sampled frame thus obtained in the B$_{435}$ band is shown in Fig.~\ref{fig:frame} (left panel), where it is also compared to a single raw exposure (right panel). As immediately visible in the figure, the software allows to reduce the pixel size to $\sim0.0125\times0.0125$ arcsec$^2$/pixel on the combined images, thus doubling the effective spatial resolution in both the U$_{220}$ and the B$_{435}$ filters.
This image processing is crucial as it allows us to resolve the stars in the  very central region of the cluster, where the stellar crowding is too severe even for the resolving power of the HRC~\citep[see also Figure 1 from][]{ha10}. 
\begin{figure*}
\centering
\includegraphics[width=11.9truecm]{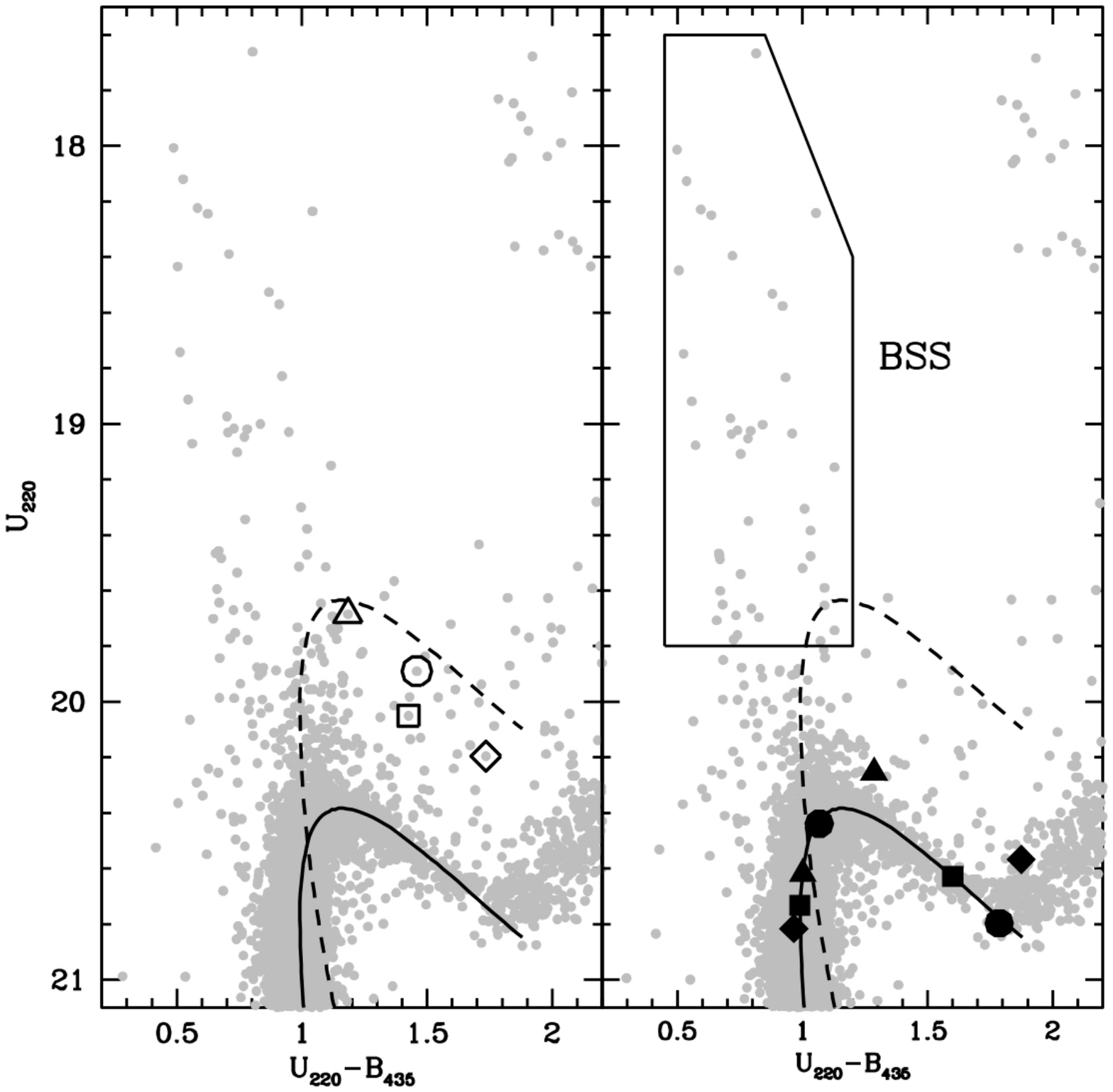}
\caption{
CMDs obtained with standard PSF fitting photometry (left panel) and our de-blending procedure with ROMAFOT (right panel). For reference, the best-fit isochrone and the equal-mass binary sequence are shown as solid and dashed lines, respectively. 
As apparent, many stars are located between these two lines in the left panel: the vast majority of these objects turned out to be blends due to poor PSF fitting. 
For illustration purposes, four blends are  highlighted  with open symbols in the left-hand panel and their corresponding de-blended components are marked with the same (solid) symbols in the right panel. Each of the four blends is the combination of a MS-TO and a SGB star. The box used to select the BSS population is also shown.
}
\label{fig:cmd_roma}
\end{figure*}
%

We first performed a standard PSF fitting photometry on the two super-sampled U$_{220}$ and B$_{435}$ frames using DAOPHOTII~\citep[][]{ste87}. The catalogue was calibrated into the VEGAMAG system using the recipe from~\citet[][]{si05} and adopting the most recent zero points available through the ACS Zeropoint Calculator\footnote{https://acszeropoints.stsci.edu}. 
All the post-MS stellar evolutionary sequences typical of a GC are well distinguishable on the first version of the CMD shown on the left panel of Fig.~\ref{fig:cmd_roma}. Still, we notice a number of objects falling in the region brighter than the Sub-Giant Branch (SGB) and bluer than the Red Giant Branch (RGB). 
\begin{figure*}
\centering
\includegraphics[width=0.6\textwidth]{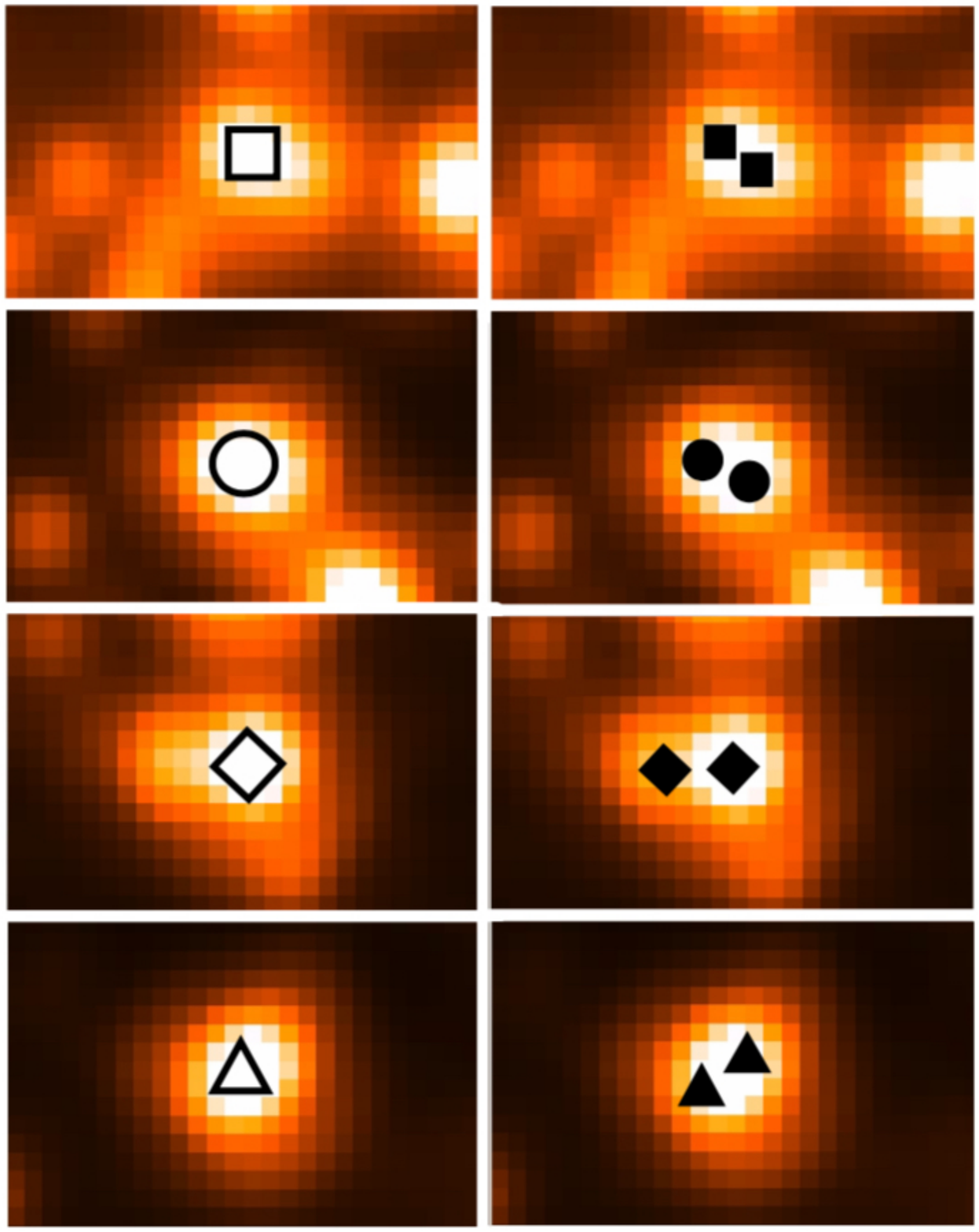}
\caption{B$_{435}$ images of the same sources discussed in Fig.~\ref{fig:cmd_roma}: the single (blended) objects in our starting photometry are marked with empty symbols in the left panels, while the corresponding resolved components, obtained after our de-blending procedure with ROMAFOT, are shown with the same (solid) symbols in the right panels.}
\label{fig:zoom_blend}
\end{figure*}
%

The origin and the reliability of these stars (sometime called ``yellow stragglers") is a well known problem and it has been discussed in many papers in the literature. The vast majority of these objects have been demonstrated to be optical blends (see, e.g., Figure 8 in \citealt{fe92} and Figure 7 in \citealt{fe91}, where a few examples of de-blending are illustrated). Here, we used the package ROMAFOT \citep{b83} to manually inspect the quality of the residuals of the initial PSF fitting process, and to iteratively improve the quality of the PSF fitting for a selection of individual problematic stars. ROMAFOT is a software developed to perform accurate photometry in crowded fields.  While the PSF fitting procedure with ROMAFOT requires a higher level of manual interaction by the user with respect to DAOPHOTII, it has the unique advantage of a graphical interface that allows the user to improve the local deconvolution of stellar profiles \citep[see also][]{mo15}. Indeed the results of the analysis with ROMAFOT fully confirm the findings by \citet{fe91,fe92}: most of the objects lying in the yellow stragglers region turned out to be blends. For the sake of illustration, in the left panel of Fig. \ref{fig:cmd_roma} we highlight four spurious sources (open symbols) due to poor PSF fitting. As shown in the right-hand panel, each of these sources is indeed the blend of a MS-TO and a SGB star (solid symbols). These sources are also highlighted with the same symbols on the B$_{435}$ images shown in Fig.~\ref{fig:zoom_blend}. In Figure \ref{fig:cmd_roma} we also plot the equal-mass binary sequence (dashed line) and the adopted BSS selection box. It is evident that the conservative BSS selection discussed in Section 3 (considering only BSSs brighter than $U=19.8$) prevents any significant contamination from unresolved binaries and blends.

\begin{figure*}
\centering
\includegraphics[width=11.9truecm]{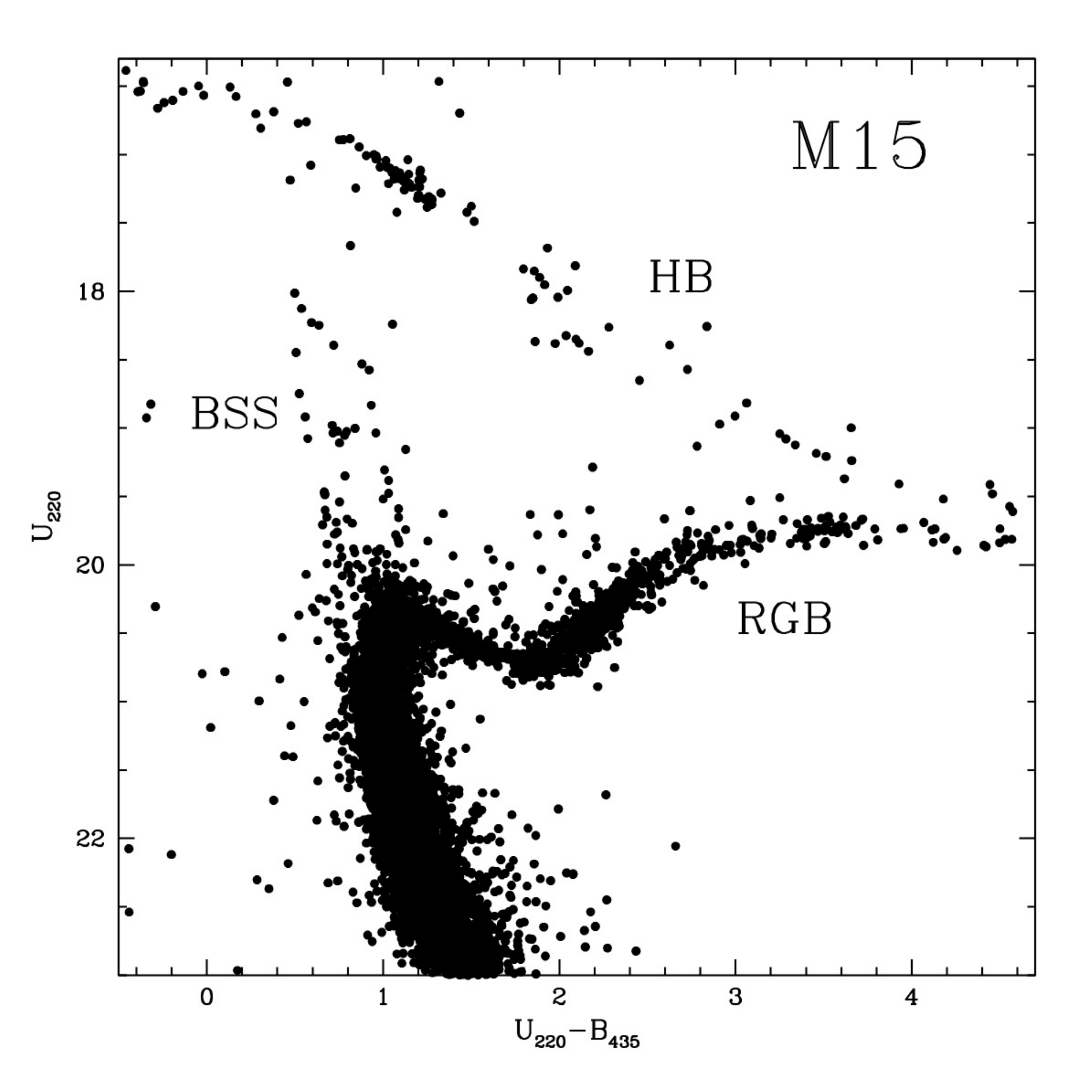}
\caption{Final, de-blended UV-CMD of the core of M15. The position in the CMD of the Blue Straggler star (BSS) population, and of the Red-Giant Branch (RGB) and Horizontal Branch (HB) evolutionary sequences is labeled.}
\label{fig:cmd_HRC}
\end{figure*}

The iterative de-blending process described above was used to optimize the PSF modeling of all the objects lying outside the well recognizable canonical evolutionary sequences in the CMD, including all the BSSs. In Fig.~\ref{fig:cmd_HRC} we show the final CMD obtained at the end of the accurate de-blending procedure. 
It should be noted that the HRC data-set that we use in this paper allows us to resolve stars with a separation larger than $\sim0.012\arcsec$, which translates into a separation of $\sim125$ au at the distance of M15 (10.4 kpc). Hence, even after our de-blending procedure, it is still possible that a few unresolved binaries and blends with separation smaller that 125 au are still populating the ``yellow straggler region". On the other hand, some of these objects could also be BSSs that are evolving from their main-sequence (core hydrogen burning) stage to a more evolved evolutionary phase (the SGB). Still, given the overall low number of BSSs, and taking into account that the SGB stage at the typical BSS mass ($\sim1.2$ M$_{\odot}$) is quite rapid, we expect just a few evolved BSSs in this part of the CMD.

\section{THE BSS POPULATION IN THE CORE OF M15}
\label{sec_bss}
The CMD shown in Fig.~\ref{fig:cmd_HRC} demonstrates the outstanding quality of the data, which allows us to properly sample the hot stellar population in the core of M15 and to detect MS stars well below the MS-TO. Thanks to the use of the U$_{220}$ filter, the hot horizontal-branch (HB) stars describe a very narrow and well defined sequence. This fact per se demonstrates already the excellent photometric quality of the data~\citep[see also][]{die07}. The same applies to the RGB stars whose sequence spans almost 2.5 mag in color on the CMD. As expected, the use of the UV filter clearly suppresses the contribution of the RGB stars to the total cluster's light, making this plane optimal for sampling the hottest stellar populations in the cluster (see also \citealt{fe97,fe03,ra17,fe18}).

\begin{figure*}
\centering
\includegraphics[width=11.9truecm]{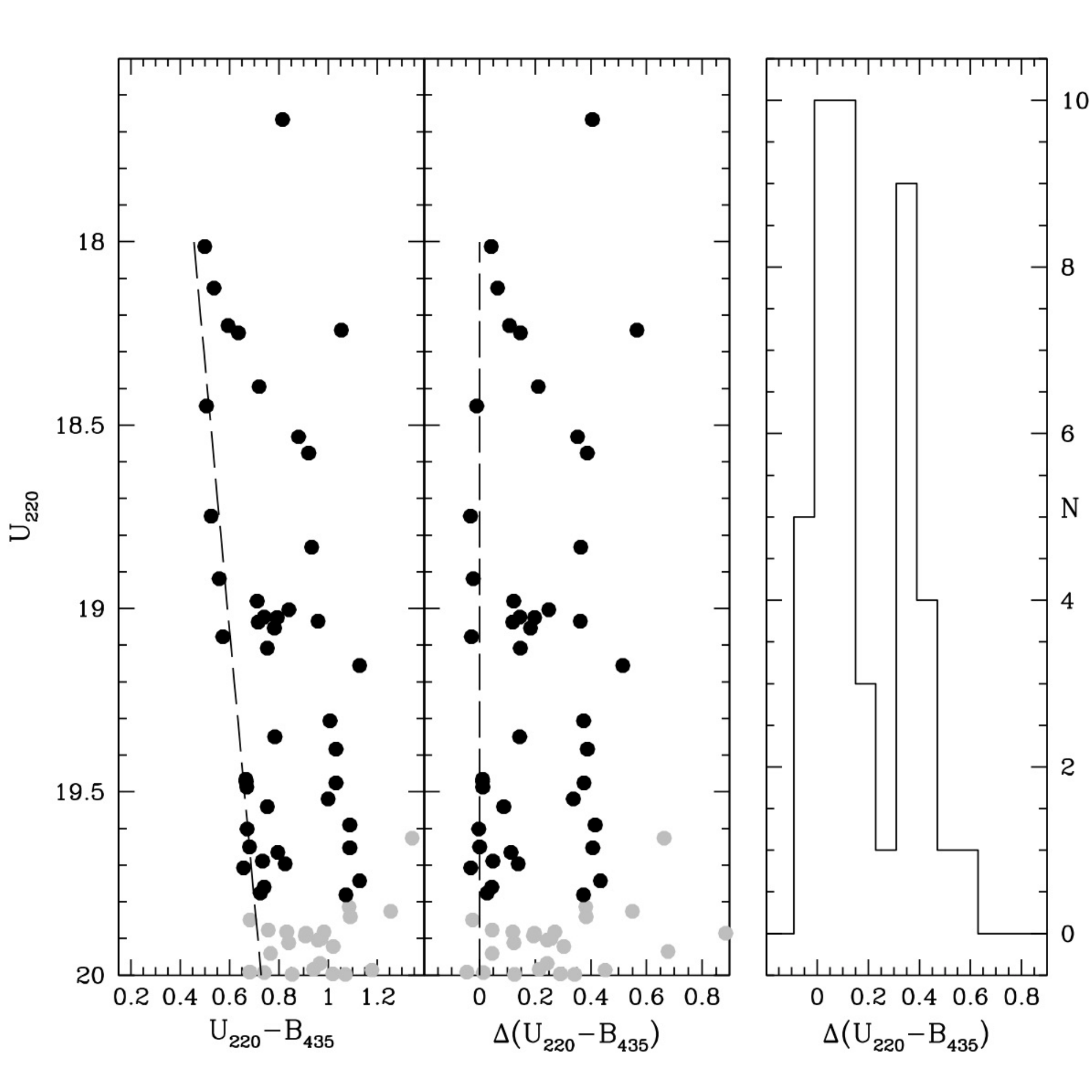}
\caption{Left Panel: Portion of the CMD zoomed on the BSS region. All the BSSs used in this study (which are brighter than U$_{220}<19.8$) are marked as black circles. The dashed line shows a fit to the bluest stars of the blue-BSS population. The line is used as reference to calculate the distribution in U$_{220}$-B$_{435}$ color of the BSS population. Central Panel: Rectification of the CMD, showing the color distance $\Delta$(U$_{220}$-B$_{435}$) of the surveyed stars from the mean ridge line of the blue-BSS population (dashed line, the same as in the left panel). Right Panel: Histogram of the BSS color distances from the mean ridge line of the blue population.}
\label{fig:bss_hist}
\end{figure*}

\subsection{A double sequence of BSS}
A surprising feature clearly visible in the CMD of Fig.~\ref{fig:cmd_HRC} is the presence of two distinct, parallel and well separated sequences of BSSs. The two sequences are almost 
vertical in the CMD, in a magnitude range $20<$U$_{220}<18$, and located at (U$_{220}$-B$_{435})\sim0.8$ and $\sim 1.2$. A similar feature was previously detected using optical bands in the GCs M30, NGC~362 and NGC~1261~\citep[][respectively]{fe09,da13,si14}. 
{\it This is the first time that such feature is detected in an UV-CMD}. 
 
\begin{figure*}
\centering
\includegraphics[width=12truecm]{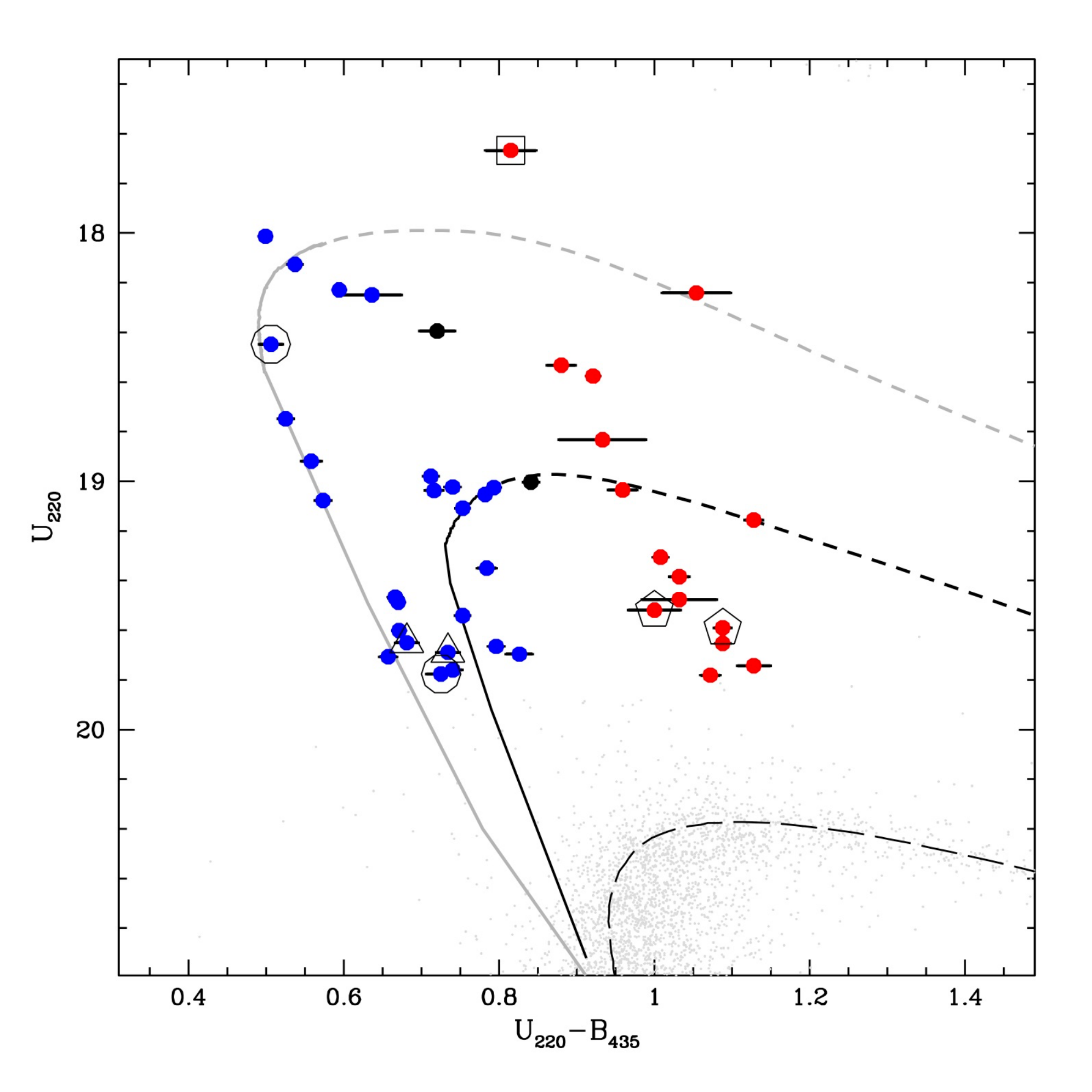}
\caption{UV-CMD zoomed on the BSS population. The stars belonging to the blue-BSS and red-BSS sequences are shown as blue and red circles, respectively. Each BSS is assigned to a given sequence according to a statistical test based on the Gaussian mixture modeling (GMM) algorithm presented in \citet{mu10}. The two stars shown as black circles have a probability of 96\% (36\%) and 69\% (31\%) to belong to the blue (red) sequence. Variable BSSs are highlighted with open symbols (see text). Two collisional isochrones, of 2 Gyr (thick grey line) and 5.5 Gyr (thick black line), are superimposed to the observed CMD. The evolutionary tracks of a $0.6+0.6 M_\odot$ and a $0.6+0.4 M_\odot$ collision product are also plotted (grey and black dashed lines, respectively). For the sake of comparison, we also mark the 12 Gyr-old isochrone for normal (single) stars (black long-dashed line).}
\label{fig:bss_zoom}
\end{figure*}

We show in Fig.~\ref{fig:bss_hist} a portion of the CMD zoomed on the BSS region.Hereafter, we will consider as bona-fide BSSs those stars populating the region of the CMD in the colour ranges explicitly mentioned above and with U$\lesssim19.8$. As discussed above, such a conservative magnitude threshold guarantees that the BSS sample is negligibly affected by residual  blended sources and unresolved binaries (see right panel in Fig. \ref{fig:cmd_roma}). On the left panel we show with a dashed line the fit to the bluest sequence of BSSs. We take this line as a reference and we then calculate the distance in (U$_{220}-$B$_{435})$ color of the BSSs observed at U$_{220}<19.8$ (black dots). In the rightmost panel we show the histogram of the distribution of the colour's distances. This is clearly not unimodal: at least two peaks, with a clear gap in-between, are well distinguishable. To assess the statistical significance of such bi-modality, we used the Gaussian mixture modeling (GMM) algorithm presented by~\citet{mu10}, which works under the assumption that the multi-modes of a distribution are well modeled via Gaussian functions. We found that the separation of the means of the 2 peaks relative to their widths is $D=4.71\pm0.78$. The parametric bootstrap probability of drawing such value of $D$ randomly from a unimodal distribution is $3.6\%$. The probability of drawing the measured kurtosis is $0.4\%$. As such, all three statistics clearly indicate that the observed BSS color distribution in the UV-CMD is bi-modal. The GMM code also provides the user with the probability distribution of each element to belong to a given peak. We find that 27 and 15 stars have a probability $>98\%$ to belong to the bluest and reddest peak, respectively. Their location in the CMD is shown in Fig. \ref{fig:bss_zoom}. The 2 stars shown as black dots in the figure have a probability of 96\% (36\%) and 69\% (31\%) to belong to the bluest (reddest) sequence. We also used the Dip test of uni-modality \citep{ha85} to further investigate on the significance of the deviation from uni-modality of the BSS color distribution shown if Fig.~\ref{fig:bss_hist}. The Dip test has the benefit of being insensitive to the assumption of Gaussianity. We found that the observed BSS color distribution has 98.5\% probability to deviate from a unimodal distribution. The  tests described above hence indicate that the bi-modality of the BSS sequence that we have found in the UV-CMD of M 15 is a statistically solid result.

In order to further prove the presence of the observed bi-modality in the BSS colour distribution, we retrieved a number of high resolution images of the core of M15 obtained with the HST/WFC3 (GO-13297; PI: Piotto). These images, obtained using the F275W and F336W filters were already used in~\citet[][]{ra17}
to investigate the dynamical age of the cluster through the study if the BSS'radial distribution~\citep[see also][]{la16}.  We stress here that the plate scale of $\sim0.04\arcsec$ offered by the WFC3 did not allow the authors to firmly detect the double BSS sequence in the core of the cluster as reported in this work. Interestingluy enough, hints for the presence of a double BSS sequence are visible in the CMD of M15 shown in Figure 22 of~\citet[][]{pio15}.

We used the photometric catalogue obtained by~\citet[][]{ra17} to resolve the BSS in a region outside the area covered by our HRC data out to a radius r=80\arcsec. Moreover we used several hundreds stars in common between the WFC3 and our HRC catalogue to accurately register the position
of the stars in the HRC catalogue to the WFC3 position. We then used ALLFRAME~\citep[][]{ste94}
to obtain accurate PSF fitting photometry of the F275W and F336W images of the WFC3 using as input catalogue of stellar centroids the coordinates of HRC stars registered on the WFC3 coordinate system. 

\begin{figure*}
\centering
\includegraphics[width=12truecm]{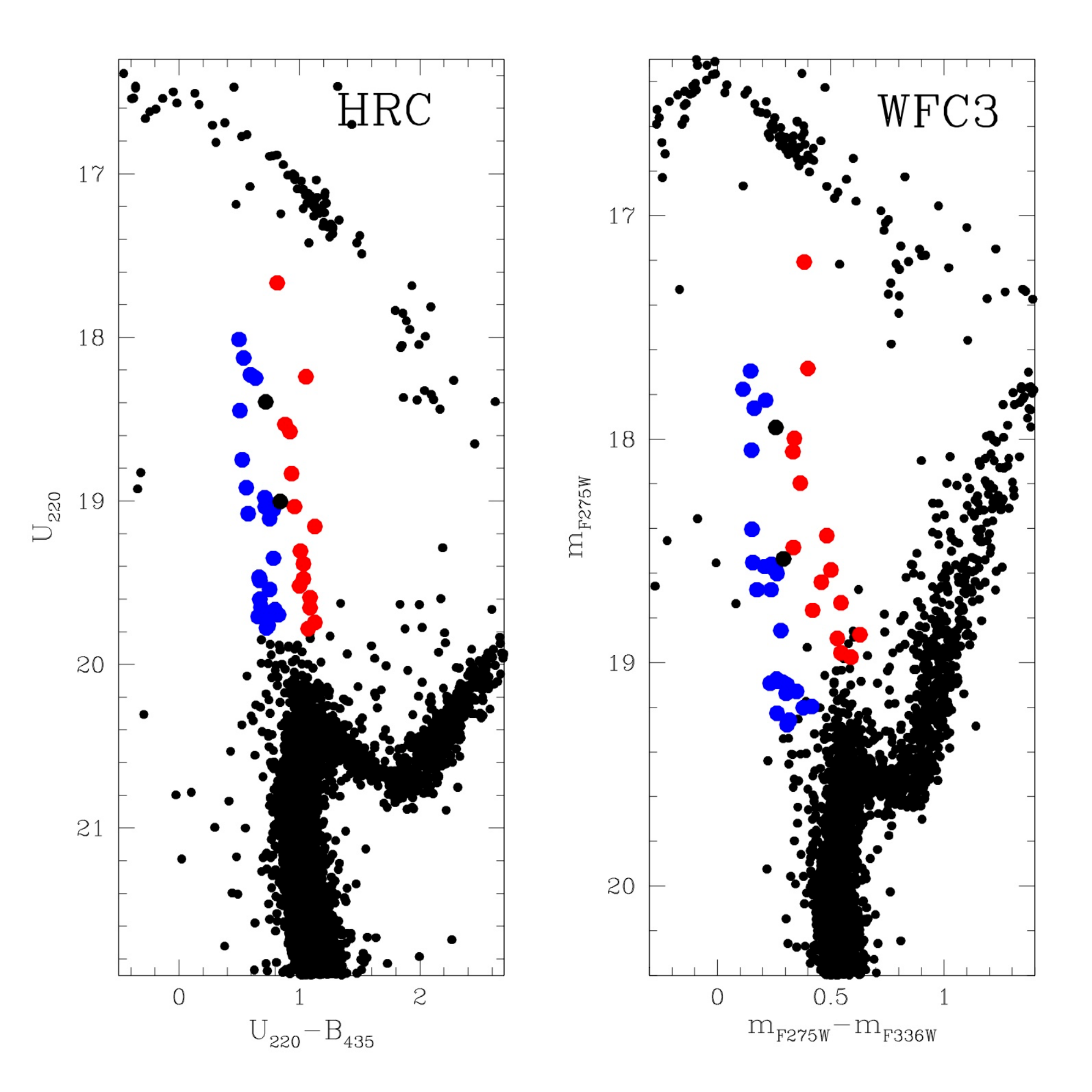}
\caption{UV-CMD obtained in the same area with the HRC (left panel) and WFC3 (right panel; see text for details) zoomed on the BSS population. The stars belonging to the blue-BSS and red-BSS sequences are shown as blue and red circles, respectively.}
\label{fig:cmd_all}
\end{figure*}

We show on Fig.~\ref{fig:cmd_all} the CMDs obtained with the two data-sets. Clearly the separation of the BSS into a blue and red sequences according to the selection done in the HRC plane, holds also in the CMD obtained with the WFC3. 

\subsection{Variable BSS}
\citet[][]{die07} used the HRC data-set to identify the variable stars in the core of M~15 (see their Table 2). We have used their catalog to identify possible variable among the BSS population. We have found 7 variables among our BSS populations, namely V20, V22, V33, V37, V39, V38, V41~\citep[nomenclature from Table 2 of][]{die07}. 
The location of these variables in the CMD is shown with open symbols in Fig. \ref{fig:bss_zoom}. Two variables (V22 and V33; open pentagons) are classified as SX-Ph, 2 as Cataclysmic Variables (CV) candidates (V39 and V41; open circles), 2 are unclassified variables observed in the CMD region of BSSs (V37 and V38; open triangles) and V20 is classified as candidate B-HB star (open square). The variables V37, V38, V39 and V41 show flickering, i.e. irregular and small (few tenths of a magnitude) variations on short timescales. For this reason \citet[][]{die07} identify them as candidate CVs. While V37 and V38 are indeed located along the BSS sequence also in the CMD shown by~\citet[][]{die07}, the authors speculate that V39 and V41 might be early CVs. Our UV-CMD convincingly shows that these stars are all BSSs. Additional observations are needed to firmly assess the nature of these variables, primarily to clarify if they are binary systems or single stars. Indeed a few W Uma Variables (which are contact binaries) have been found along  both the BSS sequences in M30 \citep{fe09} and NGC 362 \citep[][]{da13}. Using dedicated theoretical models, \citet{ji17} found that the contribution of the MT-BSSs to the blue or red sequence strongly depends on the contribution of the donor to the total luminosity of the unresolved binary system. Hence our observations further support the possibility that both collisional and MT-BSSs can populate the blue sequence. 

\subsection{A collisional sequence}

We used a set of tracks and isochrones extracted from the collisional model library of \citet{sills09} to evaluate whether the location of the blue BSS sequence in M15 is in agreement with the locus predicted by BSS collisional models (as previously found in M30 and in NGC 362). While the details of the model are described in \citet{sills09}, we list shortly here the main ingredients: (1) BSSs are formed by direct collisions between two unrelated main-sequence stars: a set of 16 cases are investigated and involving the collisions of 0.4, 0.6 and 0.8 M$_\odot$ stars. (2) Evolutionary tracks have been calculated by using the Yale Rotational Evolutionary Code (YREC, \citet{guenther92}). (3) Collisional products are generated by using the code {\it ``Make Me A Star"} (MMAS. \citet{lombardi02}) that uses the YREC results to calculate (via the so-called {\it ``sort by entropy''} method) the structure and chemical profile of the collision products. (4) Collisions are assumed to occur with a periastron separation of 0.5 times the sum of the radii of the two stars. (5) Any rotation of the collision product is neglected. (6) Finally, the recipe outlined in \citet{sills97} has been adopted to evolve the collision products from the end of the collision to the main sequence: in particular the evolution is stopped when the energy generation due to hydrogen burning was larger than that due to gravitational contraction, which corresponds to the zero age main sequence for a normal pre-main-sequence evolutionary track.

We have colored the tracks and isochrones in the U$_{220}$ and B$_{435}$ photometric bands by convolving a grid of suitable \citet{ku93} stellar spectra of appropriate metallicity ([Fe/H]$\sim-2.4$~\citealt{Ha96}, 2010 edition) with the  transmission curves of the used ACS/HRC filters. Thus, for each given stellar temperature and gravity, both the color and the bolometric corrections in the Vegamag system have been computed.
 
A 2 Gyr-old collisonal isochrone thus obtained is shown in Fig.~\ref{fig:bss_zoom} (grey thick line): a distance modulus $(m-M)_0 =15.14$ and a color excess $E(B-V)=0.08$ \citep[][2010 edition]{Ha96} have been adopted. As can be seen, the 2 Gyr isochrone well reproduces the brightest portion of the blue sequence (with U$_{220}<19.2$). To better illustrate the expected post-MS evolutionary path of collisional BSS we also plotted the track (dashed line) of a $0.6+0.6 M_\odot$ collisional product whose TO point occurs at 2 Gyr.
As sanity check we also plotted a canonical single star 12 Gyr-old isochrocrone of appropriate metallicity (black long-dashed line). As can be seen, it nicely reproduces the single star MS-TO of the cluster, thus demonstrating that the transformation and the adopted choice of distance modulus and reddening are appropriate. 
The impressive agreement of the 2 Gyr collisional isochrone with the brightest portion of the blue sequence strongly supports the hypothesis that these BSSs have been simultaneously formed by an event that, about 2 Gyr ago, led to a significant increase of the stellar interaction rate. On the other hand, the red-BSS sequence cannot be reproduced by collisional models by any of the available collisional models, regardless of the age. All the features revealed by our analysis so far nicely resemble what was found in the study of the BSS populations in M30 and NGC 362. Still, the BSSs in M15 show an additional intriguing feature.

\subsection{An additional intriguing feature}
As can be seen from Fig.~\ref{fig:bss_zoom}, a clump of 7 stars (approximately at U$_{220}=19$) together with a few sparse stars at lower luminosity, can be distinguished in the region of the CMD between the blue and red BSS sequences. As explained in Sec. \ref{sec_data}, we manually checked the accuracy of the PSF fitting solution for each of these stars. Moreover, the visual inspection of their location on the ACS/HR images indicates that they are not contaminated by bright neighbors. We also carefully analyzed the shape of the brightness profile of each star and the corresponding parameters characterizing their PSF: indeed this analysis fully confirms that they are not blends, but well-measured single stars.

Although we are dealing with a small number of stars (and the statistical significance is therefore unavoidably low), such a clump of BSSs between the two sequences has never been observed before, neither in M30 nor in NGC 362, where the overall population numbers are similar, but this region of the CMD is essentially empty. Hence, the question raises: ``what is the origin of these objects?''.
Following the scenario suggested by \citet{fe09}, stars within the gap should be evolved (post-MS) BSSs that, because of the natural stellar evolution aging process, are leaving the MS stage. 
However, for being as clumped as observed, these stars should be all evolving at the same rate and have been caught in the same evolutionary stage. This seems to be quite implausible, especially because the post-MS evolutionary time-scale is rapid. Instead the comparison with a 5.5 Gyr-old collisional isochrone (black thick line in Fig.~\ref{fig:bss_zoom}) shows an impressive match: the CMD location of the clump is nicely reproduced by the TO point of this model. Such a nice correspondence suggests that the observed clump is made of collisional BSSs with an age of 5.5 Gyr 
that are currently reaching the main sequence TO evolutionary stage. By definition, the TO is the phase during which a star is leaving the MS and moves to the SGB stage, keeping roughly the same brightness (i.e. magnitude) while becoming colder (redder). Hence it is not surprising to have an over-density at this position in the CMD. In the same plot we also show the corresponding collisional track to illustrate the post-MS evolution of the collisional BSS currently located at the MS-TO point (in this case the BSS is originated by the collision of two stars of $0.4 M_\odot$ and $0.6 M_\odot$, respectively; black dashed line).

\subsection{BSS Radial distribution}
\begin{figure*}
\centering
\includegraphics[width=11.9truecm]{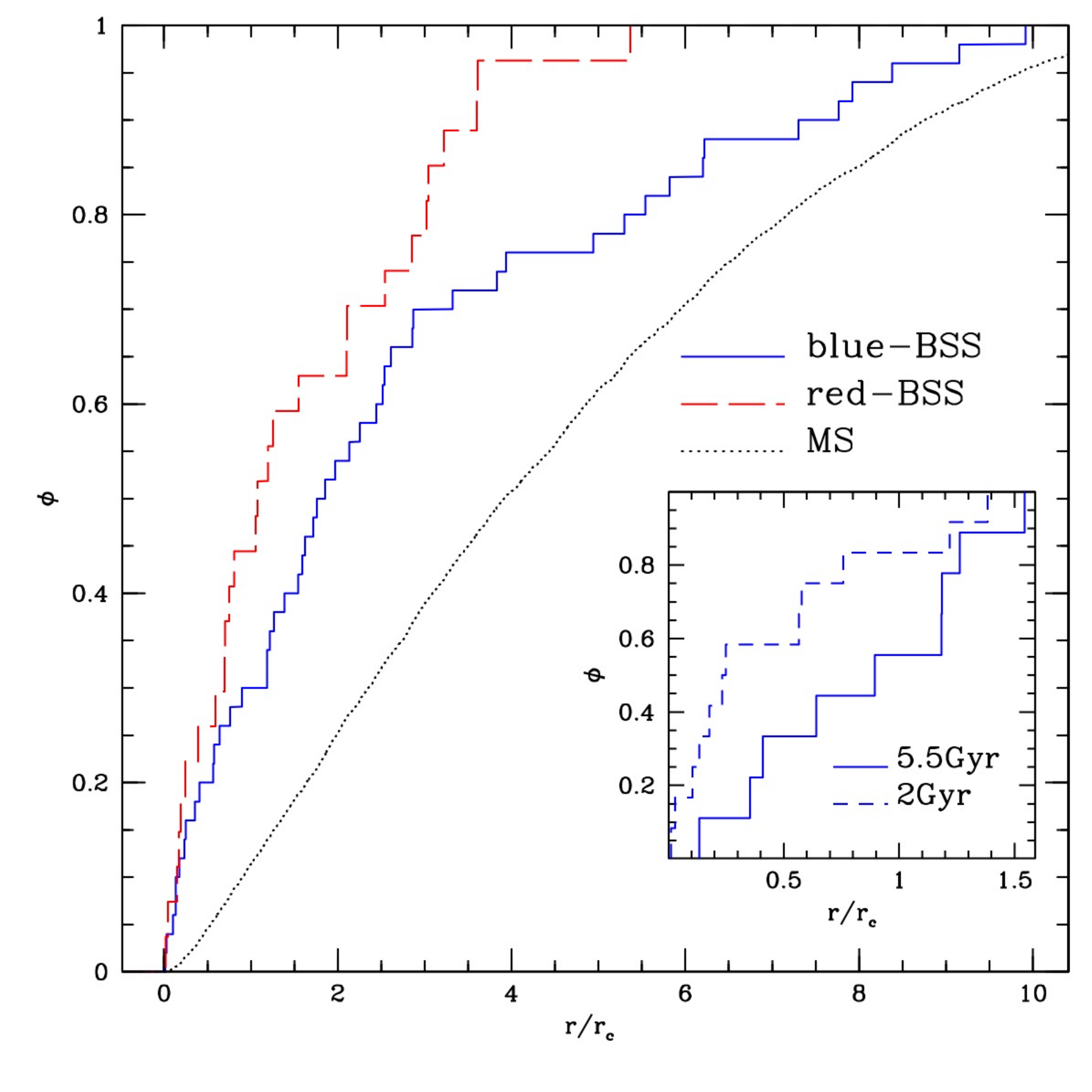}
\caption{Cumulative radial distributions of the BSSs belonging to the blue and the red populations (blue/solid and red/dashed lines, respectively) as a function of the core radius r$_c$=0\farcm14~\citealt{Ha96}, 2010 edition). The cumulative radial distribution of MS stars is also shown for comparison (black dotted line). The inset shows the cumulative radial distributions of the BSSs observed along the two branches of the blue sequence: those with CMD position well reproduced by a 2 Gyr collisional isochrone (dashed line), and those corresponding to the 5.5 Gyr collisional isochrone (solid line).}
\label{fig:radial}
\end{figure*}

We plot in Fig.~\ref{fig:radial} the cumulative radial distribution of the BSSs belonging to the blue and the red sequences (blue/solid and red/dashed lines, respectively). As explained
in Sec.~\ref{sec_bss} we used a photometric catalogue obtained with the WFC3 to extend the BSS radial distribution out to a radius r=80\arcsec. As expected the BSSs are significantly more segregated than normal MS stars (black dotted line), regardless of which population they belong to. As also found in the GCs M30 and NGC 362, the red sequence appears to be more segregated than the blue one. As discussed in~\citet[][]{da13}, this difference may offer further support for a difference formation history of the two populations. In short the blue-BSSs, born as a consequence of increase of collisions during the core-collapse, have been also kicked outward during collisional interactions while most of the red BSSs sank into the cluster center because of dynamical friction and did 
not suffer significant hardening during the core-collapse phase. The Kolgomorov-Smirnov test applied to the cumulative radial distributions suggests that the blue and the red sequences do not belong to the same parental population at only 1.5 $\sigma$ level of significance. We stress here that this is likely due to the small number statistics of the two populations.
The inset in the figure shows the cumulative radial distributions of the BSS along the two branches of the blue sequence. The BSSs populating the youngest collisional sequence (dashed line) appear more centrally segregated than the oldest ones (solid line). Although the two sub-populations include only a few stars each, the statistical significance of such a difference turns out to be of the order of 1.5-2 $\sigma$.  

\begin{figure*}
\centering
\includegraphics[width=11.9truecm]{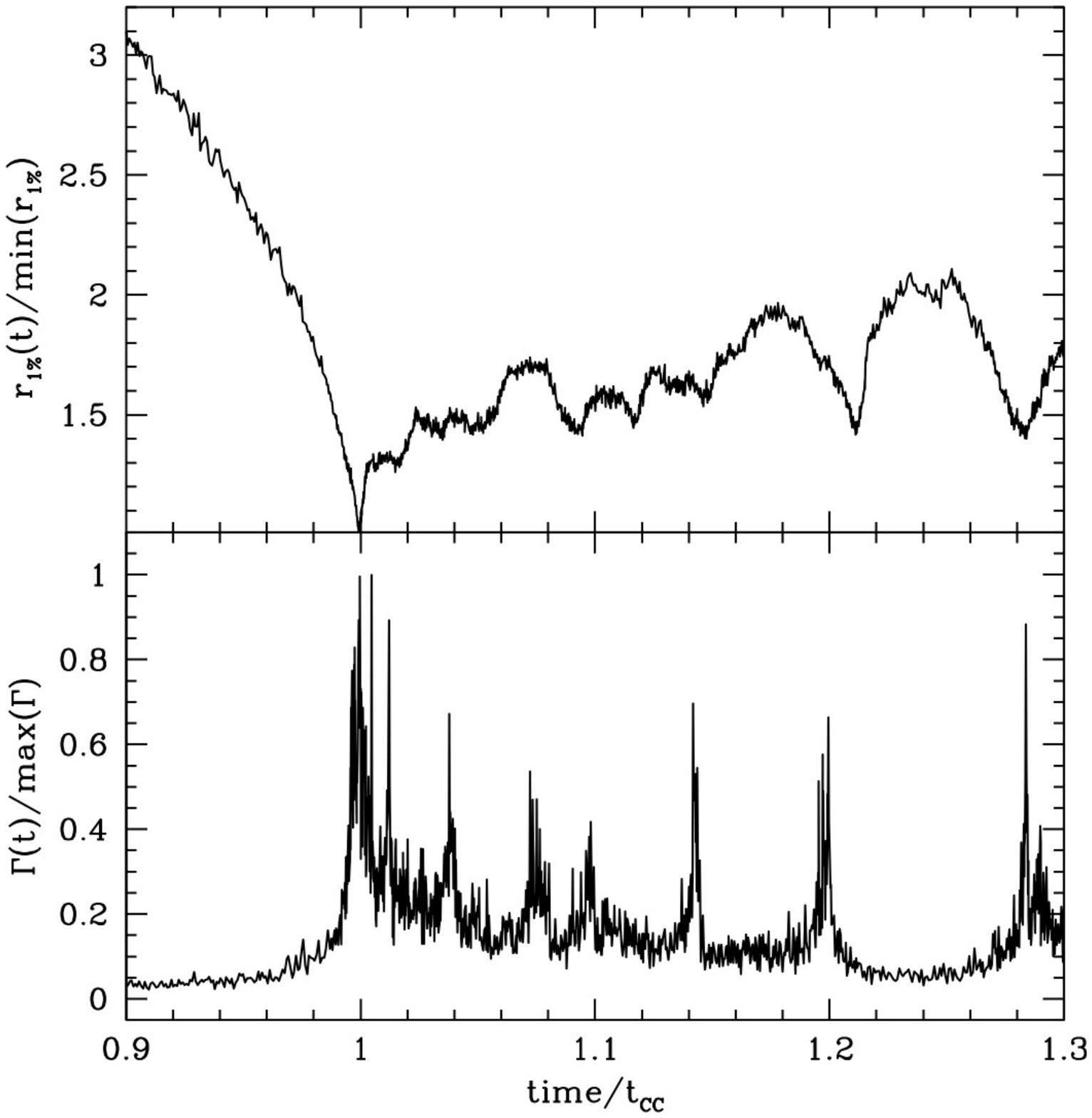}
\caption{Time evolution of the 1\% Lagrangian radius (r$_{1\%}$, top panel) and of the collisional parameter ($\Gamma$, bottom panel) as measured in a a Monte Carlo simulation of GC evolution (see text). Time is normalized to the core collapse time (t$_{CC}$ ), while r$_{1\%}$ and $\Gamma$ are normalized, respectively, to their minimum and maximum values (note that for the simulation shown here t$_{CC}\backsimeq11.9$ Gyr and the time interval shown in the plots corresponds to about 4.8 Gyr).}
\label{fig:mod}
\end{figure*}

\section{DISCUSSION}

As previously discussed, the presence of a double sequence of BSSs is a feature that was already discovered using optical photometry in the core of the GCs M30, NGC 362 and NGC 1261. 
 
\citet{fe09} first suggested that the presence of two parallel sequences of BSSs in the CMD can be explained by the coexistence of BSSs formed through two different formation mechanisms: the blue sequence would be originated mainly by collisions, while the red sequence derives from a continuous MT formation process (see also \citealt{da13} and \citealt{xi15}).
This scenario has been recently confirmed by a set of simulations that follow the evolution of a sample of BSSs generated by stellar collisions \citep{pz19}. This work concludes the blue BSS chain detected in M30 can be  indeed populated by collisional BSSs originated (over a short timescale) during the CC phase, thus finally proving that CC can be at the origin of a narrow sequence of nearly coeval Blue BSSs. 
Conversely, the photometric properties of BSSs generated by MT processes in binary systems are currently controversial, since different models seem to predict different MT-BSS distributions in the CMD (see the case of the models computed for M30 by \citealt{xi15} and \citealt{ji17}). 
In addition, even admitting that MT-BSSs may extend to the blue side of the BSS distribution in the CMD, it is unclear if and how MT processes alone could produce two split sequences, distinctly separated by gap.
Since at the moment there are no consolidated and specific models of BSS formation through MT for M15, here we focus on the blue sequence that, in this cluster, shows the most intriguing features. 
The nice match between the blue-BSS locus in the CMD and the collisional isochrones of \citet{sills09} strongly suggests that this population could be produced during a dynamical phase characterized by a high rate of stellar collisions. Moreover, in striking similarity with previous cases, the fact that the blue BSSs in the CMD appear so nicely aligned along the collisional isochrone MS suggest that they were mostly generated over a relatively short timescale (within a few $10^8$ Myrs), thus representing an essentially coeval population. The superb photometric quality of the data presented here allows us to make a significant step forward in our understanding of this phenomenon. It is particularly interesting that in M15 we are able to distinguish possible sub-structures along the collisional sequence. The two branches discovered here, in fact, can be interpreted as two generations of collisional BSSs, possibly formed during two distinct episodes of intense dynamical interactions.
Since M15 is a core-collapsed cluster (see, e.g., \citealt{dk86,che89,mu11}), such episodes can quite possibly be associated to the cluster's structural properties at the time of core collapse and during the dynamical stages following this event.
In particular, after reaching a phase of deep core collapse, clusters may undergo a series of core oscillations driven by gravothermal effects and binary activity (see, e.g., \citealt{be84,he89,co89,ti91,ma96,me96,he09,bh12}), characterized by several distinct phases of high central density followed by more extended stages during which a cluster rebounds toward a structure with lower density and a more extended core. Interestingly, \citet[][]{gr92} performed detailed numerical simulations, including post-collapse core oscillations, able to reproduce the contemporary presence of a large core radius (13 pc) and a cusp in the stellar density profile of M 15 as revealed by early HST observations from~\citet[][]{lauer91}.

The two branches of blue-BSS sequence might be the outcome of the increased rate of collisions during two separate density peaks in the post-core collapse evolution.
In order to provide a general illustration of the possible history of the cluster's structural evolution leading to the observed blue-BSS branches, we show in Fig.~\ref{fig:mod} the time evolution of the 1\% Lagrangian radius (top panel) and of the collisional parameter (bottom panel) 
as obtained from a Monte Carlo simulation run with the MOCCA code \citep{gi08}. The collisional parameter $\Gamma=\rho_c^2r_c^3/\sigma_c$ (where $\rho_c,~r_c, \hbox{and}~\sigma_c$ are, respectively, the cluster's central density, core radius and central velocity dispersion) is often used to provide a general quantitative measure of the rate of stellar collisions in a GC (see, e.g., \citealt{davies04}). The simulation presented here starts from the initial structural properties discussed in \citet{ho17}, with $7 \times 10^5$ stars, no primordial binaries and an initial half-mass radius equal to 2 pc. We point out that the adopted simulation and initial conditions do not consist in a detailed model of M15, but just illustrate the general evolution of a cluster reaching a phase of deep core collapse and undergoing core oscillations in the post-core collapse phase.

As shown in Fig.~\ref{fig:mod}, the initial deep core collapse (clearly distinguishable at $t/t_{CC}=1$) leads to the largest increase in the value of $\Gamma$, and is followed by several distinct re-collapse episodes leading to secondary peaks of different magnitudes in the collisional parameter (all smaller than the initial peak associated to the first collapse). The time interval between the various peaks in $\Gamma$ is not constant and, for the simulation presented here, typically ranges from approximately a few hundreds million years to approximately a billion years. Although we strongly emphasize that models exploring in detail the actual BSS production in each collapse episode, along with a proper characterization of the collision events and the number of stars involved, are necessary \citep[see e.g.][]{ba16}, here we suggest that the BSS populations observed along the two branches of the blue-BSS sequence in M15 might be produced during the initial deep collapse and during one of the subsequent collapse events, one sufficiently deep to trigger a significant number of collisions and BSS production. 
Note that a similar argument was used by \citet{si14} to explain the presence of a few bright BSSs located on the blue side of the blue-BSS sequence in NGC 1261. However, this cluster does not show the typical signature of core collapse, and for this reason the authors suggest that it is currently experiencing a bounce state.

\section{Summary}
We used an exceptional set of high-resolution ACS/HRC images to explore the stellar population in the collapsed core of the GC M15. A high-precision UV-CMD has revealed the existence of two clear-cut sequences of BSSs. 
In particular, we discovered that the blue sequence, which should be populated by collisional BSSs in the interpretative scenario proposed by \citet{fe09}, consists of two distinct branches nicely reproduced by collisional isochrones of 2 and 5.5 Gyr.
We interpret these features as the observational signature of two major collisional episodes 
suffered by M15, likely connected to the collapse of its core: the first one (possibly tracing the beginning of the core-collapse process) occurred approximately 5.5 Gyr ago, while the most recent one dates back 2 Gyr ago. This result reinforces the evidence of the strong link existing between the observational properties of the BSS population in a cluster and its dynamical history \citep{fe09,fe12,fe18}, and it opens a window to gather a deeper understanding of core collapse and post-core collapse evolution, as well as the link between these dynamical phases and a cluster's stellar content.

 .  

\acknowledgments
We are grateful to an anonymous referee for precious comments and suggestions.  We are grateful to Jay Anderson and Nathalie C. Haurberg for providing us with the processed ACS/HRC images. This research used the facilities of the Canadian Astronomy Data Centre operated by the National Research Council  of Canada with the support of the Canadian Space Agency. FRF is grateful to the Indiana University for the hospitality during his stay in Bloomington, when part of this work was performed.



\end{document}